\documentstyle[11pt]{article}

\def\noi{\noindent}

\makeatletter
\renewcommand{\section}{\@startsection{section}{1}{0pt}%
        {-3.5ex plus -1ex minus -.2ex}{2.3ex plus .2ex}%
        {\large\bf\protect\raggedright}}

\renewcommand{\subsection}{\@startsection{subsection}{2}{0pt}%
        {-3ex plus -1ex minus -.2ex}{1.4ex plus .2ex}%
        {\normalsize\bf\protect\raggedright}}

\renewcommand{\@oddhead}{\raisebox{0pt}[\headheight][0pt]{%
   \vbox{\hbox to\textwidth{\rightmark \hfil \rm \thepage \strut}\hrule}}}
\renewcommand{\@evenhead}{\raisebox{0pt}[\headheight][0pt]{%
   \vbox{\hbox to\textwidth{\thepage \hfil \leftmark \strut}\hrule}}}
\newcommand{\Acknow}[1]{\subsection*{Acknowledgement} #1}
\makeatother

\newcommand{\sect}[1]{Sec.\,#1}

\def\nqq{\hspace{-2em}}
\def\nhq{\hspace{-0.5em}}

\def\cm{\hspace{1cm}}
\def\inch{\hspace{1in}}

\def\eqdef{\stackrel{\rm def}=}

\def\eq{Eq.\,}
\def\eqs{Eqs.\,}
\def\beq{\begin{equation}}
\def\eeq{\end{equation}}
\def\bear{\begin{eqnarray}}
\def\al{&\nhq}
\def\lal{&&\nqq {}}               
\def\bearr{\begin{eqnarray} \lal}
\def\ear{\end{eqnarray}}
\def\earn{\nonumber \end{eqnarray}}
\def\dst{\displaystyle}
\def\tst{\textstyle}
\newcommand{\fracd}[2]{{\dst\frac{#1}{#2}}}

\def\nnn{\nonumber\\ \lal }

\def\yy{\\[5pt]}

\def\eql{\al =\al}

\def\e{{\,\rm e}}

\def\sign{\mathop{\rm sign}\nolimits}

\def\const{{\rm const}}

\def\half{{\tst\frac{1}{2}}}

\newcommand{\vars}[1]{\left\{\begin{array}{ll}#1\end{array}\right.}

\def\eps{\varepsilon}
\def\umx{u_{\max}}
\def\sph{spherically symmetric\ }

\def\bhs{black holes}

\def\df{\delta\varphi}
\def\da{\delta\alpha}
\def\db{\delta\beta}
\def\dg{\delta\gamma}
\def\pn{perturbation}
\def\pns{perturbations}

\begin{document}

\begin{center}
{\Large\bf STRUCTURE AND STABILITY OF COLD   \yy
	      SCALAR-TENSOR BLACK HOLES}        \yy
\bigskip

{\bf K.A. Bronnikov\footnote{e-mail: kb@cce.ufes.br;
permanent address: Centre for Gravitation and Fundamental Metrology,
VNIIMS, 3-1 M. Ulyanovoy St., Moscow 117313, Russia, e-mail:
kb@goga.mainet.msk.su},
G. Cl\'ement\footnote{e-mail: gecl@ccr.jussieu.fr; permanent address:
Laboratoire de Gravitation et Cosmologie Relativistes, Universit\'e Pierre
et Marie Curie, CNRS/URA769, Tour 22-12, Bo\^{\i}te 142, 4 Place Jussieu,
75252 Paris Cedex 05, France.},
C.P. Constantinidis\footnote{e-mail: clisthen@cce.ufes.br}
and J.C. Fabris\footnote{e-mail: fabris@cce.ufes.br}}
\medskip

Departamento de F\'{\i}sica, Universidade Federal do Esp\'{\i}rito Santo,
Vit\'oria, Esp\'{\i}rito Santo, Brazil
\medskip

\end{center}

\begin{abstract}
We study the structure and stability of the recently discussed \sph
Brans-Dicke black-hole type solutions with an infinite horizon area and zero 
Hawking temperature,
existing for negative values of the coupling constant $\omega$. These solutions
split into two classes: B1, whose horizon is
reached by an infalling particle in a finite proper time, and B2,
for which this proper time is infinite. Class B1 metrics are shown to be
extendable beyond the horizon only for discrete values of mass and scalar
charge, depending on two integers $m$ and $n$. In the case of even
$m-n$ the space-time is globally regular; for odd $m$ the metric changes its
signature at the horizon. All \sph solutions of the Brans-Dicke theory 
with $\omega<-3/2$ 
are shown to be linearly stable against \sph \pns. This result
extends to the generic case of the Bergmann-Wagoner class of scalar-tensor
theories of gravity with the coupling function $\omega(\phi) < -3/2$.
\end{abstract}


\section{Introduction}

    In the recent years there has been a renewed interest in scalar-tensor
    theories (STT) of gravity as viable alternatives to general relativity
    (GR), mostly in connection with their possible role in the early
    Universe. Another aspect of interest in STT is the possible existence of
    black holes (BHs) different from those well-known in GR.
    Thus, Campanelli and Lousto \cite{lousto} pointed out among the
    static, \sph solutions of the Brans-Dicke (BD) theory
    a subfamily possessing all BH properties, but (i) existing only for
    negative values of the coupling constant $\omega$ and (ii) with
    horizons of infinite area (the so-called Type B BHs \cite{we}). These
    authors argued that large negative $\omega$ are compatible with modern
    observations and that such BHs may be of astrophysical relevance.

    In Ref.\,\cite{we} it was shown, in the framework of a general
    (Bergmann-Wagoner) class of STT, that nontrivial BH solutions can exist
    for the coupling function $\omega(\phi)+3/2 <0$, and that only in 
    exceptional cases these BHs have a finite horizon area. In the BD theory
    ($\omega = \const$) such BHs were indicated explicitly; they have
    infinite horizon areas and zero Hawking temperature (``cold BHs''), thus
    confirming the conclusions of \cite{lousto}.  These BHs in turn split
    into two subclasses: B1, where horizons are attained by infalling
    particles in a finite proper time, and B2, for which this proper time is
    infinite.

    The static region of a Type B2 BH is geodesically complete since its
    horizon is infinitely remote and forms a second spatial asymptotic in a
    nonstatic frame of reference. For Type B1 BHs the global picture is
    more complex and is studied here in some detail. It turns out that the
    horizon is generically singular due to violation of analyticity, despite
    the vanishing curvature invariants. Only a discrete set of
    B1-solutions, parametrized by two integers $m$ and $n$, admits a
    Kruskal-like extension, and, depending on their parity, four different
    global structures are distinguished. Two of them, where $m-n$ is even,
    are globally regular, in two others the region beyond the horizon
    contains a spacelike or timelike singularity. 

    We also discuss the stability of STT solutions under \sph \pns. Under
    reasonable boundary conditions, it turns out that the BD solutions with
    $\omega<-3/2$ are linearly stable, and this result extends to similar
    solutions of the general STT provided the scalar field does not create
    new singularities in the static domain. For the case $\omega > -3/2$
    the stability conclusion depends on the boundary condition at a naked
    singularity, which is hard to formulate unambiguously.

    Thus some vacuum STT solutions describe regular stable BH configurations
    with peculiar global structures.

\section{Black holes in scalar-tensor theories}           

     The Lagrangian of the general (Bergmann-Wagoner) class of
     STT of gravity in four dimensions is
\beq                                               \label{L1}
     L = \sqrt{-g}\biggr[\phi R
           + \frac{\omega(\phi)}{\phi}\phi_{;\rho}\phi^{;\rho}
	                     + L_{\rm m}\biggl]
\eeq
     where $\omega(\phi)$ is an arbitrary function of the scalar field
     $\phi$ and $L_{\rm m}$ is the Lagrangian of non-gravitational matter.
     This formulation (the so-called {\it Jordan conformal
	frame\/}) is commonly considered to be fundamental since just in this
	frame the matter energy-momentum tensor $T^\mu_\nu$ obeys the
	conventional conservation law $\nabla_{\alpha}T^\alpha_\mu =0$, giving
	the usual equations of motion (the so-called atomic system of
	measurements). We consider only scalar-vacuum configurations and put
	$L_{\rm m}=0$.

     The transition to the {\it Einstein conformal frame\/},
     $g_{\mu\nu} = \phi^{-1}\bar g_{\mu\nu}$, transforms \eq (\ref{L1})
	(up to a total divergence) to the form of GR with
	a minimally coupled scalar field $\varphi$,
\beq                                                  \label{L2}
     L
     = \sqrt{-\bar g}\biggr(\bar R + \eps
         {\bar g}^{\alpha\beta}\varphi_{;\alpha}\varphi_{;\beta}\biggl),
	\cm
             \eps = \sign (\omega + 3/2),
	\cm
   \frac{d\varphi}{d\phi} = \biggl|\frac{\omega + 3/2}{\phi^2}\biggr|^{1/2}.
\eeq
     The field equations are
\beq                                                         \label{eq1}
     R_{\mu\nu} = -\eps \varphi_\mu \varphi_\nu, \cm
     \nabla^\alpha \nabla_\alpha\varphi =0
\eeq
     where we have suppressed the bars marking the Einstein frame.
	The value $\eps=+1$ corresponds to normal STT, with positive energy
	density in the Einstein frame; the choice $\eps=-1$ is anomalous. The
	BD theory corresponds to the special case $\omega = \const$, so
	that $\phi = \exp (\varphi/\sqrt{|\omega+3/2|})$.

     Let us consider a \sph field system, with the metric
\beq                                                          \label{met1}
     ds_{\rm E}^2 = \e^{2\gamma}dt^2 - \e^{2\alpha}du^2
                  - \e^{2\beta}d\Omega^2,\cm
	d\Omega^2 =
			   d\theta^2 + \sin^2 \theta d\Phi^2,
\eeq
     where E stands for the Einstein frame, $u$ is the radial
     coordinate, $\alpha$, $\beta$, $\gamma$ and the field $\varphi$ are
	functions of $u$ and $t$. Up to the end of \sect 3 we will be
	restricted to static configurations.

     The general static, \sph scalar-vacuum solution of the
     theory (\ref{L1}) is given by \cite{73,k1}
\bear                                                     \label{ds1}
     ds_{\rm J}^2 \eql \frac{1}{\phi}ds_{\rm E}^2 =
     \frac{1}{\phi}
     \biggl\{\e^{-2bu}dt^2 - \frac{\e^{2bu}}{s^2(k,u)}
       \biggr[\frac{du^2}{s^2(k,u)} + d\Omega^2\biggl]
	     \biggr\},                 \cm
     s(k,u) = \vars     {
                    k^{-1}\sinh ku,  \ & k > 0, \\
                                  u, \ & k = 0, \\
                    k^{-1}\sin ku,   \ & k < 0,     }
		\\                                         \label{phi1}
     \varphi \eql Cu + \varphi_0, \cm C, \varphi_0 =\const,
\ear
     where J denotes the Jordan frame
     and $u$ is the harmonic radial coordinate in the static space-time,
     such that $\alpha(u) = 2\beta(u) + \gamma(u)$.
     The constants $b$, $k$ and $C$ (the scalar charge) are related by
\beq                                                      \label{r1}
            2k^2\sign k = 2b^2 + \eps C^2.
\eeq
     The range of $u$ is $0 < u < \umx$, where $u=0$ corresponds to spatial
	infinity, while $\umx$ may be finite or infinite depending on $k$ and
	the behaviour of $\phi(\varphi)$. In normal STT ($\eps=+1$), by
	(\ref{r1}), we have only $k > 0$, while in anomalous STT $k$ can have
	either sign.

     According to the previous studies \cite{73,k1}, all these solutions in
     normal STT have naked singularities, up to rare exceptions
     when the sphere $u=\infty$ is regular and admits an extension of the
	static coordinate chart. An example is a conformal scalar field in
	GR viewed as a special case of STT, leading to BHs with scalar charge
	\cite{70,bek}. Even when it is the case, such configurations are
	unstable due to blowing-up of the effective gravitational coupling
	\cite{78}.

     In anomalous STT ($\eps=-1$) the following cases without naked
     singularities can be found:

\medskip\noi
     {\bf 1.} $k > 0$.
     Possible event horizons have an infinite area (Type B \bhs),
     i.e. $g_{22}\to\infty$ as $r\to 2k$.
     In BD theory, after the coordinate transformation
     $\e^{-2ku} = 1 - 2k/r \equiv P(r)$
	the solution takes the form
\beq                                                         \label{ds+}
     ds_{\rm J}^2 = P^{-\xi} ds^2_{\rm E} =
                    P^{-\xi}\Bigl(P^{a }dt^2 - P^{-a }dr^2
                                 - P^{1 - a }r^2 d\Omega^2 \Bigl),
	\qquad  \phi = P^\xi
\eeq
     with the constants related by	$(2\omega+3) \xi^2 = 1-a^2$, $a=b/k$.
     The allowed range of $a$ and $\xi$, providing a
     nonsingular horizon at $r=2k$, is
\beq
	a >1, \inch     a  > \xi \geq 2- a.                 \label{range+}
\eeq

     For $\xi < 1$ particles can arrive at the horizon in a finite proper
	time and 
	may eventually (if geodesics can be extended)
	cross it, entering the BH interior
     (Type B1 BHs \cite{we}).
     When $\xi\geq 1$, the sphere $r=2k$ is infinitely far and it takes
	an infinite proper time for a particle to reach it. As in the
	same limit $g_{22}\to \infty$, this configuration (a Type
	B2 BH \cite{we}) resembles a wormhole.

\medskip\noi
     {\bf 2.} $k = 0$.
     Just as for $k>0$, in a general STT, only Type B \bhs\ are possible
     \cite{we}), with $g_{22}\to\infty$ as $u\to \infty$.
     In particular, the BD metric is
\beq
     ds^2 = \e^{-su}\biggr[\e^{-2bu}dt^2 - \frac{\e^{2b u}}{u^2}\biggr(
	\frac{du^2}{u^2} + d\Omega^2\biggl)\biggl], \cm
       	  s^2(\omega + 3/2) = -2b^2.                          \label{ds0}
\eeq
     The allowed range of the integration constants is
     $b > 0, \quad 2b > s > -2b$.
    This range is again divided into two halves: for $s>0$ we
    deal with a Type B1 BH, for $s<0$ with that of Type B2
    ($s=0$ is excluded since leads to GR).

\medskip\noi
    {\bf 3.} $k < 0$.
    In the general STT the metric (\ref{ds1}) typically describes a
    wormhole, with two flat asymptotics at $u=0$ and $u=\pi/|k|$, provided
    $\phi$ is regular between them. In some STT the sphere $\umx = \pi/|k|$
    may be an event horizon, with $\phi \sim 1/\Delta u^2$, $\Delta u
    \equiv |u - \umx| $. In this case it has a finite area and
    $\omega(\phi) + \frac{3}{2} \to  -0$ as $u\to\umx$.
    Such metrics behave near the horizon as the extreme
    Reissner-Nordstr\"om metric.
    In BD theory we have only a wormhole solution.
\medskip\noi

    For all the BH solutions mentioned, the Hawking temperature is zero.

\section{Analytic extension and causal structure of Type B1 Brans-Dicke
         black holes}

    Let us discuss possible Kruskal-like extensions of Type B1
    BH metrics (\ref{ds+}) and (\ref{ds0}) of the BD theory.

    For (\ref{ds+}), with $a > 1 > \xi > 2-a$,
    we introduce, as usual, the null coordinates $v$ and $w$:
\beq
    v = t + x, \qquad w = t - x, \cm \cm
                                  x \eqdef \int P^{-a}dr   \label{vw}
\eeq
    where $x \to \infty$ as $r \to \infty$ and $x \to -\infty$
    as $r \to 2k$. The asymptotic behaviour of
    $x$ as $r \to 2k$ ($P \to 0$) is
    $x \propto - P^{1-a}$, and in a finite
    neighbourhood of the horizon $P=0$ one can write
\beq                                                        \label{x}
    x \equiv \half(v-w) =-\half P^{1-a}f(P)\,,
\eeq
    where $f(P)$ is an analytic function of $P$, with $f(0)=4k/(a-1)$.
    Then, let us define new null coordinates $V<0$ and $W>0$
    related to $v$ and $w$ by
\beq                                                        \label{VW}
       - v = (-V)^{-n-1}\,, \cm     w = W^{-n-1}, \cm  n=\const.
\eeq
    The mixed coordinate patch $(V,w)$ is defined for $v<0$ ($t<-x$) and
    covers the whole past horizon $v=-\infty$. Similarly, the patch $(v,W)$
    is defined for $w>0$ ($t>x$) and covers the whole future horizon
    $w=+\infty$. So these patches can be used to extend the metric through
    one or the other horizon.

    Consider the future horizon. As is easily verified, a finite value
    of the metric coefficient $g_{vW}$ at $W=0$ is achieved if we take
    $n+1 = (a-1)/(1-\xi)$, which is positive for $a>1>\xi$.
    The metric (\ref{ds+}) can be written in
    the coordinates $(v,W)$ as follows:
\bear                                                        \label{ext}
    ds^2 \eql -(n+1)f^{(n+2)/(n+1)}\cdot(1-vW^{n+1})^{-(n+2)/(n+1)}dv\,dW
\nnn  \cm
      -\fracd{4k^2}{(1-P)^2}f^{-m/(n+1)}
                    \cdot(1-vW^{n+1})^{m/(n+1)} W^{-m} d\Omega^2
\ear
    where $m = (a-1+\xi)/(1-\xi)$.

    The metric (\ref{ext}) can be extended at fixed $v$ from $W>0$ to $W<0$
    only if the numbers $n+1$ and $m$ are both integers (since otherwise
    the fractional powers of negative numbers violate the analyticity). This
    leads to a discrete set of values of the integration constants $a$ and
    $\xi$:
\beq
          a = \frac{m+1}{m-n}, \cm    \xi = \frac{m-n-1}{m-n}.   \label{qu}
\eeq
    where, according to the regularity conditions (\ref{range+}),
    $m > n \geq 0$. Excluding the Schwarzschild
    case $m=n+1$, ($\xi =0$), we see
    that regular BD BHs correspond to integers $m$ and $n$ such that
\beq
    m-2 \ge n \ge 0.                                            \label{mn}
\eeq

    The extension through the past horizon can be performed in the coordinates
    $(V,w)$ in a similar way and with the same results.

    It follows that, although the curvature scalars
    vanish on the Killing horizon $P=0$, the metric
    cannot be extended beyond it unless the constants $a$ and $\xi$
    obey the ``quantization condition'' (\ref{qu})
    and is generically singular. The Killing horizon, which
    is at a finite affine distance, is part of the boundary of the
    space-time, i.e. geodesics terminate there. A similar property was
    found in a (2+1)--dimensional model with exact power--law metric
    functions \cite{sigma}.

The $k = 0$ solution (\ref{ds0}) of the BD theory also has a Killing
horizon ($u \to \infty$) at finite geodesic distance if $s > 0$. However,
this space-time does not admit a Kruskal--like
extension and so is singular. The reason is that in this case the relation
giving the tortoise--like coordinate $x$,
\beq
x = \int\frac{{\rm e}^{2bu}}{u^2}\,du = \frac{{\rm e}^{2bu}}{2bu^2}f(u)
\eeq
(with $f(\infty) = 1$) cannot be inverted near $u = \infty$ to obtain $u$
as an analytic function of $x$.

    To study the geometry beyond the horizons of the metric (\ref{ds+}), or
    (\ref{ext}), let us define the new radial coordinate $\rho$ by
\beq                                                         \label{rho}
    P \equiv {\rm e}^{-2ku} \equiv 1 - \frac{2k}{r} \equiv \rho^{m-n}.
\eeq
    The resulting solution (\ref{ds+}), defined in the
    static region I ($P>0$, $\rho > 0$), is
\beq                                                \label{global}
    ds^2 = \rho^{n+2}\,dt^2 -
              \frac{4k^2(m-n)^2}{(1-P)^4}\,\rho^{-n-2}\,d\rho^2 -
        \frac{4k^2}{(1-P)^2}\,\rho^{-m}\,d\Omega^2\,,
                                                       \cm
    \phi = \rho^{m-n-1}.
\eeq
    By (\ref{x}), $\rho$ is related to the mixed null coordinates
    $(v,W)$ by
\beq
      \rho (v,W) = W\,[f(P)]^{1/(n+1)}[1-vW^{n+1}]^{-1/(n+1)}.
\eeq
    This relation and a similar one giving $\rho (V,w)$
    show that when the future (past) horizon
    is crossed, $\rho$ varies smoothly, changing its sign with $W$ ($V$).
    For $\rho < 0$ the metric (\ref{global}) describes
    the space-time regions beyond the horizons. Two cases must be
    distinguished according to the parity of $m{-}n$:

\begin{figure}

\unitlength=0.5mm
\linethickness{0.4pt}

\begin{picture}(100.00,130.00)(-100,0)
\put(80.00,20.00){\line(1,1){20.00}}
\put(100.00,40.00){\line(-1,1){20.00}}
\put(80.00,60.00){\line(1,1){20.00}}
\put(100.00,80.00){\line(-1,1){20.00}}
\put(80.00,100.00){\line(1,1){10.00}}
\put(80.00,20.00){\line(-1,1){20.00}}
\put(60.00,40.00){\line(1,1){20.00}}
\put(80.00,60.00){\line(-1,1){20.00}}
\put(60.00,80.00){\line(1,1){20.00}}
\put(80.00,100.00){\line(-1,1){20.00}}
\put(60.00,40.00){\line(-1,1){20.00}}
\put(40.00,60.00){\line(1,1){20.00}}
\put(60.00,80.00){\line(-1,1){20.00}}
\put(40.00,100.00){\line(1,1){20.00}}
\put(60.00,120.00){\line(-1,1){10.00}}
\put(60.00,120.00){\line(1,1){10.00}}
\put(90.00,110.00){\line(1,1){10.00}}
\put(100.00,120.00){\line(-1,1){10.00}}
\put(60.00,40.00){\line(-1,-1){20.00}}
\put(40.00,20.00){\line(1,-1){5.00}}
\put(75.00,15.00){\line(1,1){5.00}}
\put(80.00,20.00){\line(1,-1){5.00}}
\put(80.00,80.00){\makebox(0,0)[cc]{I}}
\put(80.00,40.00){\makebox(0,0)[cc]{I}}
\put(80.00,120.00){\makebox(0,0)[cc]{I}}
\put(60.00,100.00){\makebox(0,0)[cc]{II}}
\put(60.00,60.00){\makebox(0,0)[cc]{II}}
\put(60.00,20.00){\makebox(0,0)[cc]{II}}
\end{picture}

\caption {Penrose diagram for a BH with even $m-n$}
\end{figure}
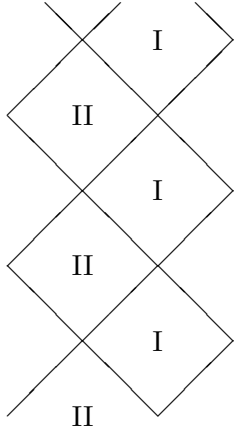

\medskip\noi
    {\bf 1.} $m-n$ is even, i.e. $P(\rho)$ is an even function.
    The two regions $\rho > 0$ and $\rho < 0$ are isometric
    ($g_{\mu\nu}(-\rho) = g_{\mu\nu}(\rho)$) for (1a) $n$ even ($m$
    even) and anti-isometric ($g_{\mu\nu}(-\rho) = -g_{\mu\nu}(\rho)$)
    for (1b) $n$ odd ($m$ odd). In this last case the metric tensor
    in region II ($\rho < 0$) has the signature ($-+++$)
    instead of ($+---$) in region I. Nevertheless, the Lorentzian nature of
    the space-time is preserved, and one can verify that all
    geodesics are continued smoothly from one region to the other (the
    geodesic equation depends only on the Christoffel symbols and is
    invariant under the anti-isometry $g_{\mu\nu} \to -g_{\mu\nu}$). The
    maximally extended space-time is regular, its Penrose diagram being the
    same in both cases 1a and 1b (Fig.\,1), an infinite
    tower of alternating regions I and II.
    However, while in case 1a timelike geodesics periodically cross the
    horizons from one region to the next, in case 1b these geodesics become
    tachyonic in region II (just as in the case of a Schwarzschild black
    hole), and generically extend to spacelike infinity ($\lambda \rightarrow
    \infty$). 
    
\medskip\noi
    {\bf 2.} $m-n$ is odd, i.e. $P(\rho)$ is an odd function;
    $P \to -\infty$ ($r \to 0$) as $\rho \to -\infty$, so that the
    metric (\ref{global}) is singular on the line $\rho = -\infty$ which is
    spacelike for $n$ odd and timelike for $n$ even. The resulting Penrose
    diagram and geodesic motion are similar to those of Schwarzschild 
    spacetime in subcase 2a ($n$ odd, $m$ even), and of extreme ($e^2 = m^2$) 
    Reissner--Nordstr\"{o}m spacetime in subcase 2b ($n$ even, $m$ odd).
 
\medskip

\section{Stability analysis}

    Let us now study small (linear) \sph \pns$\,$of the above static solutions
    (or static regions of the BHs), i.e. consider, instead of $\varphi(u)$,
\beq
	\varphi(u,t)= \varphi(u)+ \df(u,t)
\eeq
    and similarly for the metric functions $\alpha,\beta,\gamma$, where
    $\varphi (u)$, etc., are taken from the static solutions of \sect 2.
    We are working in the Einstein conformal frame and use \eqs (\ref{eq1}).

    Let us choose the coordinates in the perturbed space-time so
    that%
\footnote{The frequently used coordinate condition $\db \equiv 0$
    is here misleading since in our study configurations of utmost
    interest are type B \bhs\ and wormholes, where the area function
    $\e^\beta$ has a minimum at some value of $u$ (a throat).
    The equality $\db\equiv 0$ is there not a coordinate condition, but a
    physical restriction, forcing the throat to be at rest. It can be
    shown that the condition $\db\equiv 0$ creates spurious wall-like poles in 
    the effective potential for \pns, so that the latter exist
    separately at different sides of the throat.
}
    $\da = 2\db + \dg$, i.e. apply the same coordinate
    condition as in the metric $ds_{\rm E}$ in (\ref{ds1}).
    In this and only in this case the scalar equation
    for $\df$ decouples from other \pn\ equations
    following from (\ref{eq1}) and reads
\beq
	\e^{4\beta(u)}\delta\ddot\varphi  - \df''=0.          \label{edf}
\eeq
    Since the scalar field is the only dynamical degree of freedom, this
    equation can be used as the master one, while others only express the
    metric variables in terms of $\df$, provided the whole set of field
    equations is consistent. That it is indeed the case, can be verified
    directly: firstly, among the four different Einstein equations in
    (\ref{eq1}) only three are independent and, secondly, \eq(\ref{edf})
    may be derived from the Einstein equations. Therefore we have three
    independent equations for the three functions $\df$, $\db$ and $\dg$.

    The following stability analysis rests on \eq (\ref{edf}).
    Separating the variables,
\beq
     \df = \psi(u) \e^{i\omega t},                    \label{psi}
\eeq
    we reduce the stability problem to a boundary value problem for
    $\psi(u)$. Namely, if there is a nontrivial solution to (\ref{edf})
    with $\omega^2 <0$ under physically reasonable boundary
    conditions, then \pns\ can exponentially grow with $t$ (instability).
    Otherwise it is stable in the linear approximation.

    Suppose $-\omega^2 = \Omega^2,\ \Omega > 0$. We can
    use two forms of the radial equation (\ref{edf}): the one directly
    following from (\ref{psi}),
\beq
	\psi'' -\Omega^2 \e^{4\beta(u)}\psi=0,                \label{epsi}
\eeq
    (as before, primes denote $d/du$) and the normal Liouville
    (Schr\"odinger-like) form
\beq
	d^2 y/dx^2 - [\Omega^2+V(x)] y(x) =0,   \cm
	V(x) = \e^{-4\beta}(\beta''-\beta'{}^2).              \label{ey}
\eeq
    obtained from (\ref{epsi}) by the transformation
\beq
	\psi(u) = y(x)\e^{-\beta},\cm                         \label{tx}
				x = - \int \e^{2\beta(u)}du.
\eeq

    The boundary condition at spatial infinity ($u=0$, $x=+\infty$) is
    $\df\to 0$, or $\psi\to 0$.
    For our metric (\ref{ds1}) the effective potential $V(x)$ has the
    asymptotic form $V(x) \approx 2b/x^3$ as $x\to +\infty$,
    hence the general solution to (\ref{ey}) and (\ref{epsi}) has the
    asymptotic form
\beq
    y\sim c_1\e^{\Omega x} + c_2\e^{-\Omega x}, \cm        \label{as+}
 {\rm or} \qquad
      \psi \sim u \bigl(c_1 \e^{\Omega/u} + c_2\e^{-\Omega/u}\bigr)
\eeq
    with arbitrary constants $c_1,\ c_2$. Our boundary condition leads to
    $c_1=0$.

    For $u\to \umx$, where in many cases $\varphi \to \infty$,
    a boundary condition is not so evident. Refs.\,\cite{hod,bm} and
    others, dealing with minimally coupled or dilatonic scalar fields, used
    the minimal requirement providing the validity of the \pn\ scheme in the
    Einstein frame:
\beq
    |\df/\varphi| < \infty.                                 \label{weak}
\eeq
    In STT, where Jordan-frame and Einstein-frame metrics are related by
    $g^{\rm J}_{\mu\nu} = (1/\phi)g^{\rm E}_{\mu\nu}$,
    it seems reasonable to require that the perturbed conformal factor
    $1/\phi$ behave no worse than the unperturbed one:
\beq
    |\delta\phi/\phi| < \infty.                         \label{strong}
\eeq
    An explicit form (\ref{strong}) depends on the specific STT and
    can differ from (\ref{weak}). For example, in BD theory where $\varphi
    \propto \ln|\phi|$, (\ref{strong}) leads to 
\beq    
    |\df| < \infty.                                    \label{strongbd}
\eeq    
     We will refer to (\ref{weak}) and (\ref{strong}) (or (\ref{strongbd}))
     as the ``weak" and ``strong" boundary conditions
    respectively. If $\phi$ and $\varphi$ are regular at
    $u\to \umx$, these conditions both give $|\df|<\infty$.

    Let us discuss different cases of the STT solutions,
    assuming that the field $\phi$ is regular for $0<u<\umx$, so that the
    factor $\phi^{-1}$ in (\ref{ds1}) does not affect the range of $u$.

\medskip\noi
    {\bf 1.} $\eps=+1,\ k>0$. This is the singular solution of normal STT.
    As $u\to \infty$, $\beta(u) \sim (b-k)u$ with $b < k$, so that 
    $|x| < \infty$ and we can put $x\to 0$; the potential
    $V(x)$ has there a negative double pole, $V \sim -1/(4x^2)$, and
    (\ref{ey}) gives
\beq
    y(x) \approx \sqrt{x} (c_3 + c_4 \ln x).            \label{y1}
\eeq
    The weak boundary condition (\ref{weak}) leads to the requirement
    $|\df/\varphi| \approx |y|/u{\rm e}^{\beta} \approx
    |y|/(\sqrt{x}\ln x) < \infty$, which is met by the general asymptotic 
    solution (\ref{y1})
    and hence by its special case that joins the allowed solution
    at spatial infinity ((\ref{as+}) with  $c_1=0$).
    On the other hand, in the BD theory the strong condition (\ref{strongbd}) 
    leads to $|\psi|< \infty$ as $u\to \infty$, which
    is incompatible with the boundary condition at spatial infinity
    $\psi(u=0) = 0$, since by (\ref{epsi}) $\psi''/\psi >0$.

    We see that in this singular case the choice of a boundary
    condition is crucial for the stability conclusion. If we use the weak
    boundary condition, we conclude that the static configuration is unstable,
    in agreement with the previous work \cite{hod}. This choice of the
    weak boundary condition certainly appears appropriate in the case of GR
    with a minimally coupled scalar field, which is thus unstable. On the
    contrary, in the BD case the strong condition seems more reasonable and 
    leads to the conclusion that the BD singular solution is stable. For any
    other STT the situation must be considered separately.

\medskip\noi
    {\bf 2.} $\eps=-1,\ k>0$. This case includes some singular
    solutions and cold BHs, exemplified by (\ref{ds+}), (\ref{range+}) for
    the BD theory. Now $b > k$, so that 
    as $u\to \infty$, $x\to -\infty$ and $V(x) \approx -1/(4x^2)\to 0$.
    The general asymptotic solution to \eq (\ref{ey}) has again the form 
    (\ref{as+}).
    The weak condition (\ref{weak}) leads, as in the previous case, to the
    requirement $|y|/(\sqrt{x}\ln|x|) <\infty$ and, applied to (\ref{as+}),
    to $c_2=0$. This means that $\psi$ must tend to zero as
    both $u\to 0$ and $u\to \infty$, which is impossible due to
    $\psi''/\psi >0$. Thus the static system is stable.
    Even the weak condition is sufficient to conclude
    to stability, and a stronger one is not necessary.

\medskip\noi
    {\bf 3.} $\eps=-1,\ k=0$. There are again singular solutions and cold
    BHs. As $u \to \infty$, $x \to -\infty$ and the potential $V(x)\to 0$.
    The same reasoning as in item 2 leads to the same conclusion.

\medskip\noi
    {\bf 4.} $\eps=-1,\ k<0$. This is generically a wormhole, or in
    exceptional cases (see \sect 2) a cold BH with a finite horizon area. In
    all cases one has $x\to\-\infty$ and $V\sim 1/|x|^3$ as $u\to\umx$; even
    the weak boundary condition leads to $|\psi| < \infty$, and the
    stability is concluded just as in items 2 and 3.

\medskip
    Thus, generically, scalar-vacuum \sph solution of anomalous STT are
    linearly stable against \sph \pns.  Excluded are only the cases when the
    field $\phi$ behaves in a singular way inside the coordinate range $0 <u
    <\umx$; such cases should be studied individually.

\Acknow
{This work was partially supported by CNPq (Brazil) and CAPES (Brazil).}

\end{document}